\documentclass{article}

\usepackage{ijcai13}
\usepackage{amssymb}
\usepackage[dvips]{graphicx}
\usepackage{times}
\newtheorem{Theorem}{Theorem}
\newtheorem{Lemma}{Lemma}

\newtheorem{Example}{Example}
\newtheorem{Conjecture}{Conjecture}

\title{A Parallel Elicitation-Free Protocol for Allocating Indivisible Goods}
\author{Wei Huang$^{1}$, Jian Lou$^{1}$, \and Zhonghua Wen$^{2}$\\
1. School of Computer Science and Technology, USTC, Hefei, China\\
huangbodao@gmail.com, loujian@mail.ustc.edu.cn\\
2. College of
Information Engineering, Xiangtan
University, Xiangtan, China\\
zhonghua@xtu.edu.cn}

\begin{document}

\maketitle

\begin{abstract}
We study the problem of allocating a set of indivisible goods to
multiple agents. Recent work \cite{Bouveret} focused on allocating
goods in a sequential way, and studied what is the ``best''
sequence of agents to pick objects based on utilitarian or
egalitarian criterion. In this paper, we propose a parallel
elicitation-free protocol for allocating indivisible goods. In
every round of the allocation process, some agents will be
selected (according to some policy) to report their preferred
objects among those that remain, and every reported object will be
allocated randomly to an agent reporting it. Empirical comparison
between the parallel protocol (applying a simple selection policy)
and the sequential protocol (applying the optimal sequence)
reveals that our proposed protocol is promising. We also address
strategical issues.
\end{abstract}

\section{Introduction}
How to allocate resources among multiple agents in an efficient,
effective, and fair way is one of the most important
sustainability problems. Recently it has become an emerging
research topic in AI. Many centralized approaches to allocating
indivisible goods have been proposed (e.g., in \cite{Cramton}). In
these approaches, agents are required to fully reveal their
preferences to some central authority (which computes the final
allocation) and pay for the resources allocated to them at some
prices. However, there are some drawbacks and limitations of these
approaches:
\begin{itemize}
\item the elicitation process and the winner determination
algorithm can be very expensive;

\item agents have to reveal their full preferences, which they
might be reluctant to do (sometimes an elicitation process is
unwelcome);

\item in many real world situations (e.g., assigning courses to
students \cite{Kalinowski,Budish}, and providing employment
training opportunities to unemployed), resources must be allocated
free and monetary side payments \cite{Chevaleyre1} are impossible
or unwelcome.
\end{itemize}
So it is important to design a decentralized elicitation-free
protocol for allocating indivisible goods. \cite{Brams1} adapted a
cake-cutting protocol (a typical decentralized approach for the
allocation of divisible goods \cite{Chen}) to the allocation of
indivisible goods. However, the protocol is typically designed for
the cases when there are only two agents. \cite{Bouveret} studied
a sequential elicitation-free protocol. By applying this protocol,
any number of objects can be allocated to any number of agents.
The sequential protocol is parameterized by a sequential policy
(i.e., a sequence of agents). Agents take turns to pick objects
according to the sequence when the allocation process begins.

In this paper, we define and study a parallel elicitation-free
protocol for allocating indivisible goods to multiple agents.
According to this protocol, a parallel policy (i.e., an agent
selection policy) has to be defined before the public allocation
process can begin. At each stage of the allocation process, some
agents will be selected (according to the parallel policy) to
publicly report their preferred objects among those that remain,
and every reported object will be allocated to an agent reporting
it. If an object is reported by more than one agent, then the
agents reporting it draw lots and the winner could get it. We give
a general definition of parallel policies, which can consider the
allocation history that had happened; and provide eight different
criteria to measure the social welfare induced by parallel
policies.

In fact, any sequential policy applied in the sequential protocol
is in a specific class of parallel policies that are sensitive to
identities. The social welfare criteria considered in
\cite{Bouveret} and \cite{Kalinowski} are three of the eight
criteria proposed in our paper. We introduce two simple parallel
policies (i.e., $\varpi_A$ and $\varpi_L$), which are insensitive
to identities; and compare $\varpi_A$ and the optimal sequential
policies (for small numbers of objects and agents) with respect to
the three social welfare criteria. The results show that the
parallel protocol is promising because $\varpi_A$ outperforms the
optimal sequential policies in most cases.

We further consider strategical issues under $\varpi_A$. We show
that an agent who knows the preferences of other agents can find
in polynomial time whether she has a strategy for getting a given
set of objects regardless of uncertainty arising from lottery. We
also show that if the scoring function of the manipulator is
lexicographic, computing an optimal strategy in the sense of
pessimism is polynomial.

The remainder of this paper is structured as follows: Section 2
briefly reviews the basics of the sequential protocol. Section 3
presents the parallel protocol and introduces the two specific
parallel policies (i.e., $\varpi_A$ and $\varpi_L$). Section 4
compares $\varpi_A$ and sequential policies with respect to
several social welfare criteria. Section 5 considers strategical
issues under $\varpi_A$. Section 6 summarizes the contributions of
this work and discusses future work.

\section{Preliminaries}
A set of $m$ indivisible objects $\mathcal{O}=\{o_1,\ldots,o_m\}$
need to be allocated free to a set of $n$ agents
$\mathcal{N}=\{1,2,\ldots,n\}$. It is supposed that $m\geq n$ and
all agents have strict preferences. $\succ_i$ denotes agent $i$'s
ordinal preference (which is a total strict order) over
$\mathcal{O}$, and $rank_i(o)\in\{1,\ldots,m\}$ denotes the rank
of object $o$ in $\succ_i$. A \emph{profile} $R$ consists of a
collection of rankings, one for each agent: $R=\langle \succ_{1},
\ldots, \succ_{n} \rangle$; $Prof(\mathcal{O},\mathcal{N})$
denotes the set of possible profiles under $\mathcal{O}$ and
$\mathcal{N}$. In the following discussion, if not specified, we
only consider full independence case, where all preference
orderings are equally probable (i.e., $Pr(R)=\frac{1}{(m!)^n}$ for
every $R\in Prof(\mathcal{O},\mathcal{N})$).

Agent $i$'s value function $u_i:2^{\mathcal{O}}\rightarrow
\mathbb{R}$ specifies her valuation $u_i(B)$ on each bundle $B$
with $u_i(\emptyset)=0$. When $B=\{o\}$, we also write $u_i(B)$ as
$u_i(o)$. For any $i\in \mathcal{N}$, $B\subseteq
2^{\mathcal{O}}$, and $o\in\mathcal{O}$, it is assumed that:
\begin{itemize}
\item $u_i$ is additive, i.e., $u_i(B)=\sum_{o'\in B}u(o')$; and

\item $u_i(o)=g(rank_i(o))$, where $g$ is a non-increasing
function from $\{1,\ldots,m\}$ to $\mathbb{R}^+$.
\end{itemize}
$g$ is called the scoring function. $g$ is convex if
$g(x)-g(x+1)\geq g(y)-g(y+1)$ holds for any $x\leq y$. In this
paper, we focus on two prototypical convex scoring functions (let
$k\in\{1,\ldots,m\}$): (\emph{Borda}) $g_B(k)=m-k+1$, and
(\emph{lexicographic}) $g_L(k)=2^{m-k}$.

In the sequential protocol, agents take turns to pick objects
according to a \emph{sequential policy} $\pi\in\mathcal{N}^m$.
$\pi(i)$ denotes the $i^{th}$ agent designated by $\pi$. Given
$\pi$ and a profile $R=\langle \succ_{1}, \ldots, \succ_{n}
\rangle$, if all the agents act truthfully, then the corresponding
allocation history $h_{R}^{\pi}$ is
$\langle\pi(1),o'_1\rangle,\ldots,\langle\pi(m),o'_m\rangle$
(i.e., agent $\pi(k)$ picks object $o'_k$ at time $k$), where
$o'_k\in \mathcal{O}\setminus\{o'_l|1\leq l<k\}$ and
$o'_k\succ_{\pi(k)}o$ for every $o\in
\mathcal{O}\setminus\{o'_l|1\leq l\leq k\}$. Given a scoring
function $g$, agent $i$'s utility at $\pi$ and $R$ (i.e.,
$u_i(\pi, R)$) and $i$'s expected utility at $\pi$ (i.e.,
$u_i^*(\pi)$) are:
$$u_i(\pi, R)=\sum_{o\in\mathcal{O}_i} g(rank_i(o))$$ where
$\mathcal{O}_i=\{o'_k|1\leq k\leq m\;s.t.\;\pi(k)=i\}$, and
$$u_i^*(\pi)=\frac{\sum_{R\in Prof(\mathcal{O},\mathcal{N})}u_i(\pi, R)}{(m!)^n}$$

Given an aggregation function $F$ (which is a symmetric,
non-decreasing function from $(\mathbb{R}^+)^n$ to
$\mathbb{R}^+$), the expected social welfare of a sequential
policy $\pi$ is defined as:
$$sw_F^*(\pi)=F(u_1^*(\pi),\ldots,u_n^*(\pi)).$$
Sequential policy $\pi$ is optimal for
$\langle\mathcal{O,N},g,F\rangle$ if $sw_F^*(\pi)\geq
sw_F^*(\pi')$ for every $\pi'\in \mathcal{N}^m$.

\cite{Bouveret} considered two typical aggregation functions which
correspond to the utilitarian criterion $F_u(u_1,\ldots,
u_n)=\sum_{i=1}^n u_i$ and the Rawlsian egalitarian criterion
$F_e(u_1,\ldots, u_n)=\min\{u_i|1\leq i\leq n\}$. They also showed
that, strict alternation (i.e., $12\ldots n12\ldots n\ldots$) is
optimal for $\langle\mathcal{O},\mathcal{N},g_B,F_u\rangle$ when
$m\leq 12$ and $n=2$, and $m\leq 10$ and $n=3$. But they did not
know whether this is true for every $m$ and $n$.

The following example is modified from the one given in
\cite{Bouveret}. It illustrates the notions introduced in this
section and will be used throughout the paper.
\begin{Example}
\begin{small}
Let $m=5$, $n=3$, and $\pi=12332$. Then $\langle
u_1^*(\pi),u_2^*(\pi),u_3^*(\pi)\rangle$ is $\langle
5,7.2,7.5\rangle$ under $g_B$, and $\langle 16,17.8667,17\rangle$
under $g_L$. Consequently, $sw_{F_u}^*(\pi)=19.7$ under $g_B$,
$sw_{F_e}^*(\pi)=16$ under $g_L$, etc.

Suppose $R=\langle\succ_1,\succ_2,\succ_3\rangle$ s.t.
$\succ_1=o_1\succ o_2\succ o_3\succ o_4\succ o_5$,
$\succ_2=o_4\succ o_2\succ o_5\succ o_1\succ o_3$, and
$\succ_3=o_1\succ o_3\succ o_5\succ o_4\succ o_2$. Then
$h^\pi_R=\langle 1,o_1\rangle\langle 2,o_4\rangle\langle
3,o_3\rangle\langle 3,o_5\rangle\langle 2,o_2\rangle$. $\langle
u_1(\pi,R),u_2(\pi,R),u_3(\pi,R)\rangle$ is $\langle 5,9,7\rangle$
under $g_B$, and $\langle 16,24,12\rangle$ under $g_L$.
\end{small}
\end{Example}

\section{Parallel Protocol and Policies}
Now we introduce a parallel protocol for allocating indivisible
goods. At each stage $t$ of the allocating process, there is a
designated set of agents $\mathcal{N}_t\subseteq\mathcal{N}$ s.t.
each $i\in\mathcal{N}_t$ reports an object (her
preferred object among those that remain). If object $o$ is
reported by only one agent then it is allocated to the agent,
otherwise the agents demanding $o$ draw lots \footnote{We suppose
the lot is fair, i.e., if there are $k$ agents drawing lots then
each one of these agents has $1/k$ chance of winning the lot.} for
the right to get $o$.

The protocol is parameterized by a parallel policy. Formally, a
parallel policy is a function $\varpi: (2^{\mathcal{N}} \times
2^{\mathcal{N}})^*\rightarrow 2^{\mathcal{N}}$. Given a finite
sequence $\sigma=\langle
\mathcal{N}_1,\mathcal{N'}_1\rangle,\ldots,\langle\mathcal{N}_k,\mathcal{N'}_k\rangle$
(where for every $1\leq l\leq k$, $\mathcal{N}_l$ is the set of
agents reporting at stage $l$, and
$\mathcal{N}'_l\subset\mathcal{N}_l$ is the set of agents losing
some lottery at stage $l$), $\varpi$ designates the set of agents
reporting at stage $k+1$. An allocation history induced by
$\varpi$ is in the form of $\langle\mathcal{O}_1,
D_1\rangle\mathcal{N}'_1\langle\mathcal{O}_2,
D_2\rangle\mathcal{N}'_2\ldots\langle\mathcal{O}_p,
D_p\rangle\mathcal{N}'_p\;\textsc{Stop}$, where (suppose $1\leq
k\leq p$, and $1\leq l<p$):
\begin{itemize}
\item $\mathcal{O}_1=\mathcal{O}$,
$\mathcal{N}_1=\varpi(\epsilon)$\footnote{$\epsilon$ denotes the
empty sequence.};

\item $D_k:\mathcal{N}_k\rightarrow\mathcal{O}_k$,
$\mathcal{O}'_k=\{o\in\mathcal{O}_k|\exists
i\in\mathcal{N}_k.D_k(i)=o\}$,
$\mathcal{N}'_k\subset\mathcal{N}_k$ s.t. $\forall
o\in\mathcal{O}'_k|\{i\in\mathcal{N}_k\setminus\mathcal{N}'_k|D_k(i)=o\}|=1$;

\item $\mathcal{O}_{l+1}=\mathcal{O}_l\setminus\mathcal{O}'_l$,
$\mathcal{N}_{l+1}=\varpi(\langle\mathcal{N}_1,\mathcal{N}'_1\rangle,\ldots,\langle\mathcal{N}_l,\mathcal{N}'_l\rangle)$;

\item $\mathcal{O}_k\neq\emptyset$,
$\emptyset\subset\mathcal{N}_k\subseteq\mathcal{N}$, and
$\mathcal{O}_{p}=\mathcal{O}'_{p}$.
\end{itemize}
Intuitively, at stage $k$, $\mathcal{O}_k$ is the set of objects
remaining, $\mathcal{O}'_k$ is the set of objects reported by some
$i\in\mathcal{N}_k$, and for every $i\in\mathcal{N}_k$, $D_k(i)$
is the object reported by $i$. $\langle\mathcal{O}_k,D_k\rangle$
is called the \emph{demand situation} at $k$, and $\textsc{Stop}$
is called the \emph{termination situation}.

Given a parallel policy $\varpi$ and a profile $R=\langle
\succ_{1}, \ldots, \succ_{n} \rangle$, if all the agents act
truthfully, the set of possible histories can be represented as an
allocation structure $\mathrm{S}_R^\varpi=\langle
\mathcal{V,E}\rangle$ s.t. $\mathcal{V}$ and $\mathcal{E}$ are the
minimal sets satisfying the following rules:
\begin{itemize}
\item $\langle
\mathcal{O},D:\varpi(\epsilon)\rightarrow\mathcal{O}\rangle\in\mathcal{V}$
s.t. $rank_i(D(i))=1$ for every $i\in\varpi(\epsilon)$;

\item if there exists a history $h=\ldots\langle\mathcal{O}_k,
D_k\rangle\mathcal{N}'_k$ $\langle\mathcal{O}_{k+1},
D_{k+1}\rangle\ldots$ induced by $\varpi$ such that:

\begin{itemize}
\item$\langle\mathcal{O}_k, D_k\rangle\in\mathcal{V}$, and

\item$\forall i\in\mathcal{N}_{k+1}\forall
o\in\mathcal{O}_{k+1}\setminus\{D_{k+1}(i)\}.D_{k+1}(i)\succ_i o$,
\end{itemize}
then $\langle\langle\mathcal{O}_k,
D_k\rangle,\mathcal{N}'_k,\langle\mathcal{O}_{k+1},
D_{k+1}\rangle\rangle\in\mathcal{E}$, and
$\langle\mathcal{O}_{k+1}, D_{k+1}\rangle\in\mathcal{V}$;

\item$\textsc{Stop}\in\mathcal{V}$, and if there exists a history
$h=\ldots\langle\mathcal{O}_k,
D_k\rangle\mathcal{N}'_k\;\textsc{Stop}$ induced by $\varpi$ s.t.
$\langle\mathcal{O}_k, D_k\rangle\in\mathcal{V}$ then
$\langle\langle\mathcal{O}_k,
D_k\rangle,\mathcal{N}'_k,\textsc{Stop}\rangle\in\mathcal{E}$.
\end{itemize}
It is easy to find that $\mathrm{S}_R^\varpi$ is acyclic, and
$\langle\mathcal{O},D\rangle$ is the root.

Since the allocation process from some demand situation $v\in
\mathcal{V}$ is nondeterministic in general, each rational agent
$i$ is often concerned with her expected utility $\hat{u_i}(v)$
and the minimal utility $\underline{u_i}(v)$ that she can get
regardless of uncertainty. Formally, given a scoring function
$g$, $\hat{u_i}(v)=\underline{u_i}(v)=0$ if $v=\textsc{Stop}$;
otherwise (suppose $v=\langle
\mathcal{O}',D':\mathcal{N}'\rightarrow\mathcal{O}'\rangle$):
$$\hat{u_i}(v)=w+\frac{\sum_{v'\in\mathcal{V}}\#E_{v'}^v\cdot
\hat{u_i}(v')}{\#out_v}$$
$$\underline{u_i}(v)=\min\{\underline{u_i}(\mathcal{N}'',v')|\langle
v,\mathcal{N}'',v'\rangle\in\mathcal{E}\},\;\;\textrm{where}$$
\begin{itemize}
\item
$w=\frac{g(rank_i(D'(i)))}{|\{j\in\mathcal{N}'|D'(j)=D'(i)\}|}$ if
$i\in\mathcal{N}'$, $w=0$ otherwise;

\item $\#E_{v'}^v=|\{\mathcal{N}''\subset \mathcal{N}|\langle
v,\mathcal{N}'',v'\rangle\in\mathcal{E}\}|$;

\item $\#out_v=|\{\langle\mathcal{N}'',v'\rangle\in
2^{\mathcal{N}}\times \mathcal{V}|\langle
v,\mathcal{N}'',v'\rangle\in\mathcal{E}\}|$;

\item $\underline{u_i}(\mathcal{N}'',v')=\underline{u_i}(v')$ if
$i\in\mathcal{N}''\cup(\mathcal{N}\setminus\mathcal{N}')$,
$\underline{u_i}(\mathcal{N}'',v')=\underline{u_i}(v')+g(rank_i(D'(i)))$
otherwise.
\end{itemize}
$\hat{u_i}(\varpi,R)=\hat{u_i}(v)$ and
$\underline{u_i}(\varpi,R)=\underline{u_i}(v)$ are called agent
$i$'s expected utility and minimum utility at $\varpi$ and $R$,
respectively, where $v$ is the root of $\mathrm{S}_R^\varpi$.

Each agent $i\in\mathcal{N}$ can evaluate a given parallel policy
$\varpi$ according to 4 values, i.e., $v_i(y,z,\varpi)$ where:
\begin{itemize}
\item $y,z\in\{\textsf{u,e}\}$,

\item $v_i(\textsf{u},z,\varpi)=\frac{\sum_{R\in
Prof(\mathcal{O},\mathcal{N})}u_i(z,\varpi, R)}{(m!)^n}$,

\item $v_i(\textsf{e},z,\varpi)=\min\{u_i(z,\varpi, R)|R\in
Prof(\mathcal{O},\mathcal{N})\}$,

\item $u_i(\textsf{u},\varpi, R)=\hat{u_i}(\varpi,R)$, and
$u_i(\textsf{e},\varpi, R)=\underline{u_i}(\varpi,R)$.
\end{itemize}
The social welfare induced by $\varpi$ (i.e., $sw(x,y,z,\varpi)$)
can be measured by the 8 possible orderings over 3 elements taken
from $\{\textsf{u,e}\}$. Formally, $x,y,z\in\{\textsf{u,e}\}$,
$sw(\textsf{u},y,z,\varpi)=\sum_{i=1}^n v_i(y,z,\varpi)$, and
$sw(\textsf{e},y,z,\varpi)=\min\{v_i(y,z,\varpi)|1\leq i\leq
n\}$\footnote{Intuitively, $\textsf{u}$ and $\textsf{e}$ denote
the utilitarian principle and the egalitarian principle in social
welfare aggregation, respectively. }.

Any sequential policy $\pi$ can be seen as a parallel policy
$\varpi_\pi$ s.t. $\varpi_\pi(\epsilon)=\pi(1)$ and
$\varpi_\pi(\sigma_k)=\pi(k+1)$ for every $1\leq k<m$, where
$\sigma_k=\langle\{\pi(1)\},\emptyset\rangle,\ldots,\langle\{\pi(k)\},\emptyset\rangle$.
For every profile $R$, there is only one possible history in
$\mathrm{S}_R^{\varpi_\pi}$. So
$\hat{u_i}(\varpi_\pi,R)=\underline{u_i}(\varpi_\pi,R)=u_i(\pi,R)$,
$v_i(\textsf{u},z,\varpi_\pi)=u_i^*(\pi)$, and
$sw(x,\textsf{u,u},\varpi_\pi)=sw_{F_x}^*(\pi)$.

In this paper, we introduce two specific parallel policies:
\emph{all--reporting} $\varpi_A$, where all the agents report at
every stage, and \emph{loser--reporting} $\varpi_L$, where all the
agents losing some lot at the current stage report at the next
stage. Formally, $\varpi_A(\sigma)=\mathcal{N}$ for any sequence
$\sigma$; $\varpi_L(\epsilon)=\mathcal{N}$, and
\begin{displaymath}
\varpi_L(\ldots,\langle\mathcal{N}_k,\mathcal{N'}_k\rangle)=\left\{\begin{array}{ll} \mathcal{N'}_k & \textrm{if} \: \mathcal{N'}_k\neq\emptyset\\
\mathcal{N} & \mathrm{otherwise}
\end{array} \right.
\end{displaymath}
$\varpi_L$ guarantees that every agent can get $\frac{m}{n}$
objects at least. So in the eyes of pessimists, it may be a better
choice than $\varpi_A$.

Note that neither $\varpi_A$ nor $\varpi_L$ mentions identities of
agents. We called this kind of parallel policies are insensitive
to identities. We can get Lemma 1 directly.
\begin{Lemma}\label{lem:lemma1}
Let parallel policy $\varpi$ be insensitive to identities. Then
for every $y,z\in\{\textsf{u,e}\}$, and $i,j\in \mathcal{N}$, we
have $v_i(y,z,\varpi)=v_j(y,z,\varpi)$, and
$sw(\textsf{u},y,z,\varpi)=n\cdot v_i(y,z,\varpi)=n\cdot
sw(\textsf{e},y,z,\varpi)$.
\end{Lemma}
\begin{figure}
\begin{center}
\includegraphics [width=75mm,height=60mm]{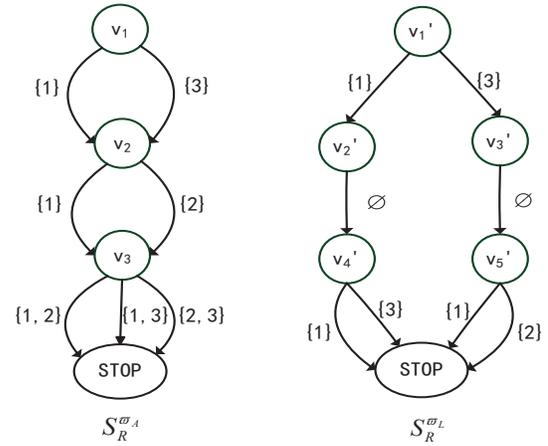}
\caption{Allocation structures of $\varpi_A$ and $\varpi_L$}
\end{center}
\end{figure}
\begin{Example}
\begin{small}
Consider the situation depicted in Example 1. Figure 1 shows the
allocation structures of $\varpi_A$ and $\varpi_L$, where (let
$1\leq p\leq 3$, $1\leq q\leq 5$, and $\textsf{ud}$ denote the
undefined value):
\begin{itemize}
\item
$v_p=\langle\mathcal{O}_p,D_p:\mathcal{N}\rightarrow\mathcal{O}_p\rangle$,
$v'_q=\langle\mathcal{O}'_q,D'_q:\mathcal{N}'_q\rightarrow\mathcal{O}'_q\rangle$;

\item $\mathcal{O}_1=\mathcal{O}'_1=\mathcal{O}$,
$\mathcal{O}_2=\mathcal{O}'_2=\mathcal{O}'_3=\{2,3,5\}$,
$\mathcal{O}_3=\{5\}$, $\mathcal{O}'_4=\{3,5\}$, and
$\mathcal{O}'_5=\{2,5\}$;

\item $\mathcal{N}'_1=\mathcal{N}'_4=\mathcal{N}'_5=\mathcal{N}$,
$\mathcal{N}'_2=\{1\}$, and $\mathcal{N}'_3=\{3\}$;

\item $d_p\in (\mathcal{O}_p)^{|\mathcal{N}|}$ s.t.
$d_p[i]=D_p(i)$ for every $i\in \mathcal{N}$, i.e., $d_1=\langle
1,4,1\rangle$, $d_2=\langle 2,2,3\rangle$, and $d_3=\langle
5,5,5\rangle$; and

\item $d'_q\in
(\mathcal{O}'_q\cup\{\textsf{ud}\})^{|\mathcal{N}|}$ s.t.
$d'_q[i]=D'_q(i)$ if $i\in \mathcal{N}'_q$, otherwise
$d'_q[i]=\textsf{ud}$, i.e., $d'_1=\langle 1,4,1\rangle$,
$d'_2=\langle 2,\textsf{nd},\textsf{nd}\rangle$, $d'_3=\langle
\textsf{nd},\textsf{nd},3\rangle$, $d'_4=\langle 3,5,3\rangle$,
and $d'_5=\langle 2,2,5\rangle$.
\end{itemize}
Under $g_B$,
$\langle\hat{u}_1(\varpi_A,R),\hat{u}_2(\varpi_A,R),\hat{u}_3(\varpi_A,R)\rangle=\langle
4.8333,8,$ $7.5\rangle$, and $sw(\textsf{u,u,u,}\varpi_A)=20.382$,
etc. Under $g_L$, $\langle \underline{u_1}(\varpi_L,R),$
$\underline{u_2}(\varpi_L,R),\underline{u_3}(\varpi_L,R)\rangle=\langle
8,16,12\rangle$, and $sw(\textsf{e,e,e,}\varpi_L)=4$.
\end{small}
\end{Example}

\section{Comparison}
\cite{Bouveret} studied what are the sequential policies
maximizing social welfare. They considered a utilitarian principle
and an egalitarian principle, in which the social welfare induced
by a sequential policy $\pi$ is measured by the values of
$sw(\textsf{u,u,u},\varpi_\pi)$ and
$sw(\textsf{e,u,u},\varpi_\pi)$, respectively. They computed the
optimal sequential policies for small numbers of objects and
agents using an exhaustive search algorithm, and further
conjectured that the problem of finding an optimal sequential
policy is \textsf{NP}-hard. It has been proved that the
alternating policy (i.e., $1212\ldots$) maximizes the value of
$sw(\textsf{u,u,u},\varpi_\pi)$ for two agents under Borda scoring
function \cite{Kalinowski1}. However, the general problem (i.e.,
finding a sequential policy maximizing the value of $sw(\textsf{u,u,u},\varpi_\pi)$ and
$sw(\textsf{e,u,u},\varpi_\pi)$, for more than two agents, or
under other scoring functions) is still open.

On another hand, parallel policy $\varpi_A$ (i.e., all--reporting)
is very natural and simple, and does not have costly procedures
like finding optimal sequence in sequential protocol. By applying
$\varpi_A$, every agent has a chance to get a remaining object in
every round of the parallel allocation process. But we don't know
if $sw(\textsf{u,u,u},\varpi_A)$ and $sw(\textsf{e,u,u},\varpi_A)$
can be computed in polynomial time. We conjecture it's much harder
than computing $sw(\textsf{u,u,u},\varpi_\pi)$ and
$sw(\textsf{e,u,u},\varpi_\pi)$, because complications arise not
only from uncertainty over profiles but from uncertainty over
lots.

In this section, we will compare the parallel protocol (applying
$\varpi_A$) and the sequential protocol (applying the optimal
sequential policies) with respect to several social welfare
criteria. We demonstrate experimentally that in most cases,
$\varpi_A$ is better than sequential policies. To sum up, the
parallel protocol is promising.

\begin{table}
\caption{$\pi^*$,
$sw(\textsf{u},\textsf{u},\textsf{u},\varpi_{\pi^*})$, and
$sw(\textsf{u},\textsf{u},\textsf{u},\varpi_A)$ under
$g_B$}\label{tab:table1} \setlength\tabcolsep{3pt}
\begin{center}
\begin{scriptsize}
\begin{tabular}{|c|c|c|c|c|c|c|c|c|c|}
\hline
 & \multicolumn{3}{|c|}{$n=2$} & \multicolumn{3}{|c|}{$n=3$} & \multicolumn{3}{|c|}{$n=4$}\\
\hline
$m$ & $\pi^*$ & $sw_{\pi^*}$ & $sw_A$ & $\pi^*$ & $sw_{\pi^*}$ & $sw_A$ & $\pi^*$ & $sw_{\pi^*}$ & $sw_A$ \\
\hline
$4$ & \underline{2}\underline{2} & $12.292$  & $12.292$ & \underline{3}1 & $13.083$ & $13.297$ & \underline{4} & $13.583$ & $13.885$ \\
\hline
$5$ & \underline{2}\underline{2}1 & $18.625$ & $18.625$  &  \underline{32} &  $20.033$ & $20.382$ & \underline{4}1 & $20.800$ & $21.351$ \\
\hline
$6$ & \underline{2}\underline{2}\underline{2} & $26.396$  & $26.396$ &  \underline{3}\underline{3} & $28.622$ &  $28.840$  & \underline{4}\underline{2} & $29.600$ & $30.377$ \\
\hline
$7$ & \underline{2}\underline{2}\underline{2}1 & $35.396$  & $35.396$ &  \underline{3}\underline{3}1 & $38.511$ &  $38.864$ & & & \\
\hline
$8$ & \underline{2}\underline{2}\underline{2}\underline{2} & $45.820$  & $45.820$ &  \underline{3}\underline{3}\underline{2} & $49.936$ &  $50.381$ & & & \\
\hline
$9$ & \underline{2}\underline{2}\underline{2}\underline{2}1 & $57.487$ & $57.487$ &  &  & & & &\\
\hline
$10$ & \underline{2}\underline{2}\underline{2}\underline{2}\underline{2} & $70.569$ & $70.569$ & & & & & &\\
\hline
\end{tabular}
\end{scriptsize}
\end{center}
\end{table}
\begin{table}
\caption{$\pi^*$,
$sw(\textsf{u},\textsf{u},\textsf{u},\varpi_{\pi^*})$, and
$sw(\textsf{u},\textsf{u},\textsf{u},\varpi_A)$ under $g_L$}
\label{tab:table2} \setlength\tabcolsep{3pt}
\begin{center}
\begin{scriptsize}
\begin{tabular}{|c|c|c|c|c|c|c|c|c|c|}
\hline
 & \multicolumn{3}{|c|}{$n=2$} & \multicolumn{3}{|c|}{$n=3$} & \multicolumn{3}{|c|}{$n=4$} \\
\hline
$m$ & $\pi^*$ & $sw_{\pi^*}$ & $sw_A$ & $\pi^*$ & $sw_{\pi^*}$ & $sw_A$ & $\pi^*$ & $sw_{\pi^*}$ & $sw_A$ \\
\hline
$4$ & \underline{2}\underline{2} & $20.458$  & $20.458$ & \underline{3}1 & $23.000$ & $23.460$ & \underline{4} & $24.417$ & $25.458$ \\
\hline
$5$ & \underline{2}\underline{2}1 & $44.725$ & $44.725$  &  \underline{3}\underline{2} &  $51.933$ & $53.028$ & \underline{4}1 & $56.350$ & $58.477$\\
\hline
$6$ & \underline{2}\underline{2}\underline{2} & $95.371$  & $95.371$ &  \underline{3}\underline{3} & $114.27$ &  $115.63$ & \underline{4}\underline{2} & $125.26$ & $129.80$ \\
\hline
$7$ & \underline{2}\underline{2}\underline{2}1 & $199.49$  & $199.49$ &  \underline{3}\underline{3}1 & $244.64$ &  $247.13$ & & &\\
\hline
$8$ & \underline{2}\underline{2}\underline{2}\underline{2} & $412.91$  & $412.91$ &  \underline{3}\underline{3}\underline{2} & $516.09$ &  $520.79$ & & &\\
\hline
$9$ & \underline{2}\underline{2}\underline{2}\underline{2}1 & $847.64$ & $847.64$ &  &  & & & & \\
\hline
$10$ & \underline{2}\underline{2}\underline{2}\underline{2}\underline{2} & $1731.0$ & $1731.0$ & & & & & &\\
\hline
\end{tabular}
\end{scriptsize}
\end{center}
\end{table}
\begin{table}
\caption{$\pi^*$,
$sw(\textsf{e},\textsf{u},\textsf{u},\varpi_{\pi^*})$, and
$sw(\textsf{e},\textsf{u},\textsf{u},\varpi_A)$ under $g_B$}
\label{tab:table3} \setlength\tabcolsep{2pt}
\begin{center}
\begin{scriptsize}
\begin{tabular}{|c|c|c|c|c|c|c|c|c|c|}
\hline
& \multicolumn{3}{|c|}{$n=2$} & \multicolumn{3}{|c|}{$n=3$} & \multicolumn{3}{|c|}{$n=4$} \\
\hline
$m$ & $\pi^*$ & $sw_{\pi^*}$ & $sw_A$ & $\pi^*$ & $sw_{\pi^*}$ & $sw_A$ & $\pi^*$ & $sw_{\pi^*}$ & $sw_A$ \\
\hline
$4$ & \underline{2}21 & $6.000$  & $6.146$ & \underline{3}3 & $3.750$ & $4.432$ & \underline{4} & $2.500$ & $3.471$\\
\hline
$5$ & 1\underline{2}22 & $9.000$ & $9.313$  &  \underline{3}32 &  $5.000$ & $6.794$ & \underline{4}4 & $4.500$ & $5.338$\\
\hline
$6$ & \underline{2}\underline{2}21 & $13.125$  & $13.198$ &  \underline{3}321 & $9.000$ &  $9.613$ & \underline{4}43 & $5.833$ & $7.594$\\
\hline
$7$ & 1\underline{2}2\underline{2}2 & $17.333$  & $17.698$ &  \underline{3}2133 & $12.250$ &  $12.955$  & & & \\
\hline
$8$ & \underline{2}2\underline{2}1\underline{2} & $22.725$  & $22.910$ &  11332232 & $15.000$ &  $16.794$ & & & \\
\hline
$9$ & 1\underline{2}\underline{2}22\underline{2} & $28.429$ & $28.744$ &  &  & & & &\\
\hline
$10$ & \underline{2}21\underline{2}\underline{2}21 & $35.200$ & $35.285$ & & & & & &\\
\hline
\end{tabular}
\end{scriptsize}
\end{center}
\end{table}
\begin{table}
\caption{$\pi^*$,
$sw(\textsf{e},\textsf{u},\textsf{u},\varpi_{\pi^*})$, and
$sw(\textsf{e},\textsf{u},\textsf{u},\varpi_A)$ under $g_L$}
\label{tab:table4} \setlength\tabcolsep{2pt}
\begin{center}
\begin{scriptsize}
\begin{tabular}{|c|c|c|c|c|c|c|c|c|c|}
\hline
& \multicolumn{3}{|c|}{$n=2$} & \multicolumn{3}{|c|}{$n=3$} & \multicolumn{3}{|c|}{$n=4$} \\
\hline
$m$ & $\pi^*$ & $sw_{\pi^*}$ & $sw_A$ & $\pi^*$ & $sw_{\pi^*}$ & $sw_A$ & $\pi^*$ & $sw_{\pi^*}$ & $sw_A$ \\
\hline
$4$ & \underline{2}21 & $10.000$  & $10.229$ & \underline{3}3 & $7.000$ & $7.820$ & \underline{4} & $3.750$ & $6.365$\\
\hline
$5$ & \underline{2}2\underline{2} & $21.667$ & $22.363$  &  \underline{3}32 &  $16.000$ & $17.676$ & \underline{4}4 & $12.400$ & $14.619$\\
\hline
$6$ & \underline{2}2\underline{2}1 & $47.500$  & $47.686$ &  \underline{3}321 & $36.533$ &  $38.543$ & \underline{4}43 & $29.333$ & $32.450$\\
\hline
$7$ & \underline{2}2\underline{2}\underline{2} & $98.400$  & $99.745$ &  \underline{3}3213 & $80.229$ &  $82.377$  & & & \\
\hline
$8$ & \underline{2}2\underline{2}\underline{2}1 & $205.40$  & $206.46$ &  \underline{3}32132 & $168.14$ &  $173.60$ & & & \\
\hline
$9$ & \underline{2}2\underline{2}\underline{2}11 & $421.59$ & $423.82$ &  &  & & & &\\
\hline
$10$ & \underline{2}2\underline{2}\underline{2}1\underline{2} & $862.79$ & $865.50$ & & & & & &\\
\hline
\end{tabular}
\end{scriptsize}
\end{center}
\end{table}

We first compare $\varpi_A$ and sequential policies with respect
to the utilitarian criterion considered by Bouveret and Lang, which are $sw(\textsf{u,u,u},\varpi_A)$
and $sw(\textsf{u,u,u},\varpi_\pi)$, respectively.
For small numbers
of objects and agents (i.e., $m$ and $n$), we compute the optimal
sequential policies (denoted by $\pi^*$) and
$sw(\textsf{u,u,u},\varpi_{\pi^*})$ by use of the tool provided by
Bouveret and Lang (http://
recherche.noiraudes.net/en/sequences.php), and compute
$sw(\textsf{u,u,u},\varpi_A)$ by use of an exhaustive method (on a
PC \textsl{Intel(R) Core(TM) i5-3570K @3.4Ghz}). The time required to compute
$sw(\textsf{u,u,u},\varpi_A)$ grows dramatically in the number n
of agents and in the number m of objects. We find that with $n=3$
and $m=8$ the computation of $sw(\textsf{u,u,u},\varpi_A)$
requires about 8 minutes, but $sw(\textsf{u,u,u},\varpi_A)$ can
not be computed in 12 hours with $n=3$ and $m=9$. The results
under Borda and Lexicographic scoring functions (i.e., $g_B$ and
$g_L$) are shown in Table \ref{tab:table1} and Table
\ref{tab:table2}\footnote{Note that in Tables 1 to 4,
$\underline{n}$, $sw_{\pi^*}$, and $sw_A$ denote the sequence
$12\ldots n$, the social welfare induced by $\pi^*$ and
$\varpi_A$.}, respectively. From Table \ref{tab:table1} and Table
\ref{tab:table2}, we can get that when $n=2$ and $10\geq m\geq 4$,
the values of $\pi^*$ and $\varpi_A$ are equal; however, when
$n>2$, the values of $\varpi_A$ are strictly greater than those of
$\pi^*$. These results suggest that, for small numbers of agents
and objects, we could have a better utilitarian social welfare if
we apply $\varpi_A$ rather than $\varpi_{\pi^*}$. We conjecture
that it is not a coincidence, but we could not find a proof.
\begin{Conjecture}\label{con:con1}
Under any convex scoring function $g$ and for any number $m\geq n$
of objects,
$sw(\textsf{u},\textsf{u},\textsf{u},\varpi_{\pi^*})=sw(\textsf{u},\textsf{u},\textsf{u},\varpi_A)$
when $n=2$, and
$sw(\textsf{u},\textsf{u},\textsf{u},\varpi_{\pi^*})<sw(\textsf{u},\textsf{u},\textsf{u},\varpi_A)$
when $n>2$.
\end{Conjecture}

We also compare $\varpi_A$ and sequential policies with respect to
the egalitarian criterion considered by Bouveret and Lang. The
results under $g_B$ and $g_L$ are shown in Table \ref{tab:table3}
and Table \ref{tab:table4}, respectively. We find that the values
of $\varpi_A$ are strictly greater than those of $\pi^*$ in all
the test cases. In fact, $\varpi_A$ is insensitive to identities.
So according to Lemma \ref{lem:lemma1},
$sw(\textsf{e},\textsf{u},\textsf{u},\varpi_A)=1/n\cdot
sw(\textsf{u},\textsf{u},\textsf{u},\varpi_A)$, which is
definitely the fairest way to divide
$sw(\textsf{u},\textsf{u},\textsf{u},\varpi_A)$. However,
sequential policies are sensitive to identities. From Conjecture
\ref{con:con1}, we further conjecture that $\varpi_A$ will always
be a better choice than any sequential policy for the balance of
utilitarianism and egalitarianism.
\begin{Conjecture}\label{con:con2}
Under any convex scoring function $g$, for any number $n$ of
agents and any number $m\geq n$ of objects,
$sw(\textsf{e},\textsf{u},\textsf{u},\varpi_{\pi^*})<sw(\textsf{e},\textsf{u},\textsf{u},\varpi_A)$.
\end{Conjecture}

\cite{Kalinowski} considered a different egalitarian principle in
which the social welfare induced by a sequential policy $\pi$ is
measured by the value of $sw(\textsf{u,e,u},\varpi_\pi)$. They
also computed the optimal sequential policies (denoted by $\pi^*$)
under $g_B$ when $n=2$ and $p\leq 8$. We compute the values of
$sw(\textsf{u},\textsf{e},\textsf{u},\varpi_{\pi^*})$ and
$sw(\textsf{u},\textsf{e},\textsf{u},\varpi_A)$ by use of an
exhaustive method. The result is shown in Table \ref{tab:table5}.
Again, $\varpi_A$ outperforms sequential policies in all the test
cases.
\begin{table}
\caption{$\pi^*$,
$sw(\textsf{u},\textsf{e},\textsf{u},\varpi_{\pi^*})$, and
$sw(\textsf{u},\textsf{e},\textsf{u},\varpi_A)$ under $g_B$ when
$n=2$} \label{tab:table5} \setlength\tabcolsep{6pt}
\begin{center}
\begin{small}
\begin{tabular}{|c|c|c|c|}
\hline
$m$ & $\pi^*$ & $sw(\textsf{u},\textsf{e},\textsf{u},\varpi_{\pi^*})$ & $sw(\textsf{u},\textsf{e},\textsf{u},\varpi_A)$\\
\hline
$2$ & 12 & $1.500$ & $1.750$ \\
\hline
$3$ & 122 & $3.000$ & $3.500$ \\
\hline
$4$ & 1221 & $5.667$ & $5.958$ \\
\hline
$5$ & 12122 & $8.483$ &  $8.992$ \\
\hline
$6$ & 121221 & $12.397$ & $12.736$ \\
\hline
$7$ & 1212122 & $16.560$ & $17.082$ \\
\hline
$8$ & 12122121 & $21.738$ & $22.129$\\
\hline
\end{tabular}
\end{small}
\end{center}
\end{table}

\section{Strategical Issues under $\varpi_A$}

In this section, we will discuss strategical issues under
all--reporting policy $\varpi_A$, which is one of the simplest
parallel policies that are insensitive to identities. As most collective decision mechanisms, $\varpi_A$ is not
\emph{strategyproof}. See Example 2. If all the agents play
sincerely, i.e., report their preferred object at each stage, then
$\hat{u}_1(\varpi_A,R)=\frac{1}{2}g(1)+\frac{1}{2}g(2)+\frac{1}{3}g(5)$
and $\underline{u_1}(\varpi_A,R)=0$. Suppose 1 is a pessimist and
believes that she cannot win any lottery. Then she is concerned
only with the utility she can get regardless of uncertainty. If 1
knows other agents' preferences and plays strategically, then she
reports $o_2$ first and she can get $g(2)$ units of utility at
least, which is better than $0=\underline{u_1}(\varpi_A,R)$.

Someone may want to study the impact of strategic behavior on the
complete-information extensive-form game of such parallel
allocation procedures \footnote{In \cite{Kalinowski2}, the
allocation procedure applying the sequential protocol, is viewed
as a finite repeated game with perfect information, where all
agents act strategically. }. However, it is supposed that the
environment matches decentralized elicitation-free protocols'
application motivation. That is to say, we suppose that it is hard
to learn self-interested agents' preferences in advance
\footnote{In the environments where every agent can learn other
agent's preferences in advance, centralized allocation methods
need to be taken into consideration instead. Because in these
cases, the prerequisite to the protection of private preferences
is ruined.}. So we accept the assumption made in \cite{Bouveret},
i.e., all agents but the only one manipulator act truthfully. In
the following discussion, without loss of generality, let 1 be the
manipulator that knows the rankings of the other agents (i.e.,
$\langle\succ_2,\ldots,\succ_n\rangle$), and $o_1\succ_1
o_2\succ_1\ldots\succ_1 o_m$.

Under $\varpi_A$, a \emph{strategy} for agent 1 is a sequence of
objects $\tau=o'_1,\ldots,o'_T$ s.t. $\forall
t,t'\in\{1,\ldots,T\}\:(o'_t\in\mathcal{O}\textrm{ and
}o'_t=o'_{t'}\textrm{ iff }t=t')$ holds. That is to say, $\tau$
specifies which object 1 should report at any stage $1\leq t\leq
T$. Some strategies may fail because some object that 1 intends to
report has already been allocated. We say strategy $\tau$ is
\emph{well--defined} with respect to
$\langle\succ_2,\ldots,\succ_n\rangle$ if at any stage
$t\in\{1,\ldots,T\}$, object $o'_t$ is still available, and there
is no object available after stage $T$.

A manipulation problem $M$ (for agent 1) consists of
$\langle\succ_2,\ldots,\succ_n\rangle$, and a target set of
objects $\mathcal{S}\subseteq\mathcal{O}$. A well-defined strategy
$\tau$ is \emph{successful} for $M$ if, assuming the agents 2 to n
act sincerely, $\tau$ ensures that agent 1 gets all objects in
$\mathcal{S}$. Solving $M$ consists in determining whether there
exists a successful strategy. Below we show that the manipulation
problem $M$ can be solved in polynomial time. First, we define
some notions: for every $i\in\mathcal{N}$,
$A,B\subseteq\mathcal{O}$ s.t. $A\cap B=\emptyset$,
$\textsc{Better}_i(A,B)=\{o\in A|(\forall o'\in B)o\succ_i o'\}$,
and $\textsc{Best}_i(A)=o\in A$ s.t. $o\succ_i o'$ for every
$o'\in A\setminus\{o\}$. We can get Lemma 2 directly.
\begin{Lemma}
Let $A\subseteq C\subseteq\mathcal{O}$, $B\subseteq
D\subseteq\mathcal{O}$, and $C\cap D=\emptyset$. Then for any
$i\in\mathcal{N}$,
$\textsc{Better}_i(A,D)\subseteq\textsc{Better}_i(C,D)\subseteq\textsc{Better}_i(C,B)$,
and if $\textsc{Best}_i(C)\in A$ then
$\textsc{Best}_i(C)=\textsc{Best}_i(A)$, otherwise
$\textsc{Best}_i(C)\succ_i\textsc{Best}_i(A)$.
\end{Lemma}

Second, for a target set $\mathcal{S}\subseteq\mathcal{O}$, we
construct a sequence
$\rho_{\mathcal{S}}=\langle\mathcal{O}'_1,\mathcal{S}_1,\mathcal{O}_1\rangle,\langle\mathcal{O}'_2,\mathcal{S}_2,\mathcal{O}_2\rangle,\ldots$
as follows:
\begin{small}
\begin{itemize}
\item $\mathcal{O}'_1=\mathcal{O}$, $\mathcal{N}'=\{2,\ldots,n\}$;

\item $\mathcal{S}_k=\bigcup_{i\in
\mathcal{N}'}\textsc{Better}_i(\mathcal{O}'_k\cap
\mathcal{S},\mathcal{O}'_k\setminus\mathcal{S})$;

\item
$\mathcal{O}_k=\{o\in\mathcal{O}'_k\setminus\mathcal{S}|(\exists
i\in\mathcal{N}')o=\textsc{Best}_i(\mathcal{O}'_k\setminus\mathcal{S})\}$;

\item
$\mathcal{O}'_{k+1}=\mathcal{O}'_k\setminus(\mathcal{S}_k\cup\mathcal{O}_k)$.
\end{itemize}
\end{small}
Obviously, for every $o\in\mathcal{O}$, there exists one and only
$k\geq 1$ s.t. $o\in \mathcal{S}_k\cup \mathcal{O}_k$. We denote
by $\textsf{app}_{\mathcal{S}}(o)$ the number $k$.
\begin{Lemma}
Let $\mathcal{S}'\subseteq \mathcal{S}\subseteq\mathcal{O}$,
$\rho_{\mathcal{S}'}=\langle\mathcal{O}'_1,\mathcal{S}_1,\mathcal{O}_1\rangle,\langle\mathcal{O}'_2,\mathcal{S}_2,\mathcal{O}_2\rangle,\ldots$,
and
$\rho_{\mathcal{S}}=\langle\mathcal{O}''_1,\mathcal{S}^*_1,\mathcal{O}^*_1\rangle,\langle\mathcal{O}''_2,\mathcal{S}^*_2,\mathcal{O}^*_2\rangle,\ldots$.
Then for every $k\geq 1$ we have $\bigcup_{t\geq 1}^k
\mathcal{S}_{t}\subseteq\bigcup_{t\geq 1}^k \mathcal{S}^*_{t}$ and
$\bigcup_{t\geq
1}^k(\mathcal{S}_t\cup\mathcal{O}_t)\subseteq\bigcup_{t\geq
1}^k(\mathcal{S}^*_t\cup\mathcal{O}^*_t)$.
\end{Lemma}
\textsc{Proof.} (Sketch)
\begin{small}
According to Lemma 2, we have
$\mathcal{S}_1\subseteq\mathcal{S}^*_1$ and
$\mathcal{S}_1\cup\mathcal{O}_1\subseteq\mathcal{S}^*_1\cup\mathcal{O}^*_1$.
Now assume that $\bigcup_{t\geq 1}^k
\mathcal{S}_{t}\subseteq\bigcup_{t\geq 1}^k \mathcal{S}^*_{t}$ and
$\bigcup_{t\geq
1}^k(\mathcal{S}_t\cup\mathcal{O}_t)\subseteq\bigcup_{t\geq
1}^k(\mathcal{S}^*_t\cup\mathcal{O}^*_t)$ for any $k<p$. Then
$\mathcal{O}'_p\supseteq \mathcal{O}''_p$. Let
$\mathcal{N}'=\{2,\ldots,n\}$.
\begin{enumerate}
\item According Lemma 2, $\mathcal{S}_p=\bigcup_{i\in
\mathcal{N}'}\textsc{Better}_i(\mathcal{O}'_p\cap
\mathcal{S}',\mathcal{O}'_p\setminus\mathcal{S}')\subseteq\bigcup_{i\in
\mathcal{N}'}\textsc{Better}_i(\mathcal{O}'_p\cap
\mathcal{S},\mathcal{O}''_p\setminus\mathcal{S})$. Pick an object
$o$ from $\mathcal{S}_p$. If $o\in \mathcal{O}''_p\cap\mathcal{S}$
then there must be $i\in \mathcal{N}'$ s.t.
$o\in\textsc{Better}_i(\mathcal{O}''_p\cap
\mathcal{S},\mathcal{O}''_p\setminus\mathcal{S})$, i.e., $o\in
\mathcal{S}_p^*$. Otherwise
$o\in(\mathcal{O}'_p\setminus\mathcal{O}''_p)\cap\mathcal{S}$,
i.e., there must be some $q<p$ s.t. $o\in\mathcal{S}_q^*$. So
according to the assumption, we have $\bigcup_{t\geq 1}^p
\mathcal{S}_{t}\subseteq\bigcup_{t\geq 1}^p \mathcal{S}^*_{t}$.

\item Pick an object $o'$ from $\mathcal{O}_p$. Then there exists
$i\in \mathcal{N}'$ s.t.
$o'=\textsc{Best}_i(\mathcal{O}'_p\setminus\mathcal{S}')$. It is
easy to find that
$\mathcal{O}''_p\setminus\mathcal{S}\subseteq\mathcal{O}'_p\setminus\mathcal{S}'$.
So if $o'\in \mathcal{O}''_p\setminus\mathcal{S}$ then
$o'=\textsc{Best}_i(\mathcal{O}''_p\setminus\mathcal{S})$, i.e.,
$o'\in\mathcal{O}^*_p$. If $o'\in\mathcal{S}$ then there exists
some $q\leq p$ s.t. $o'\in\mathcal{S}_q^*$. Otherwise (i.e.,
$o'\in \mathcal{O}'_p\setminus(\mathcal{O}''_p\cup\mathcal{S})$)
there exists some $q'<p$ s.t. $o'\in\mathcal{O}_{q'}^*$.
\end{enumerate}
According to items 1 and 2 and the assumption, we have
$\bigcup_{t\geq 1}^p \mathcal{S}_{t}\subseteq\bigcup_{t\geq
1}^p\mathcal{S}^*_{t}$ and $\bigcup_{t\geq
1}^p(\mathcal{S}_t\cup\mathcal{O}_t)\subseteq\bigcup_{t\geq
1}^p(\mathcal{S}^*_t\cup\mathcal{O}^*_t)$. $\quad\square$
\end{small}

Now we can give a simple characterization of successful strategies
in manipulation problems.
\begin{Theorem}
Let
$M=\langle\langle\succ_2,\ldots,\succ_n\rangle,\mathcal{S}\rangle$
be a manipulation problem, and
$\rho_{\mathcal{S}}=\langle\mathcal{O}''_1,\mathcal{S}^*_1,\mathcal{O}^*_1\rangle,\langle\mathcal{O}''_2,\mathcal{S}^*_2,\mathcal{O}^*_2\rangle,\ldots$.
There exists a successful strategy for $M$ iff for any $k\geq 1$
we have $k>|\bigcup_{1\leq t\leq k}\mathcal{S}^*_{t}|$. Moreover,
in this case any strategy $\tau$ starting by reporting the objects
in $\mathcal{S}$, and reporting $o$ before stage
$\textsf{app}_{\mathcal{S}}(o)$ for every $o\in \mathcal{S}$, (and
completed so as to be well-defined) is successful.
\end{Theorem}
\textsc{Proof.} (Sketch)
\begin{small}
We prove the statement by induction on the size of the target set
$\mathcal{S}$. In the case when $\mathcal{S}$ is a singleton
\{o\}, it is easy to find that $|\bigcup_{1\leq t\leq
k}\mathcal{S}^*_{t}|\leq|\{o\}|=1$ for every $k\geq 1$. So
$\mathcal{S}^*_{1}=\emptyset$ (i.e.,
$\textsf{app}_{\mathcal{S}}(o)>1$) iff $k>|\bigcup_{1\leq t\leq
k}\mathcal{S}^*_{t}|$ for any $k\geq 1$. If
$\mathcal{S}^*_1=\emptyset$ (i.e., no agent in $\{2,\ldots,n\}$
reports $o$ at stage 1) then 1 can get $o$ by reporting $o$ first.
If $\mathcal{S}^*_1\neq\emptyset$ (i.e., there must be some agent
in $\{2,\ldots,n\}$ reporting $o$ at stage 1) then there exists no
successful strategy for $M$.

Now assume that the statement holds for any target set whose size
is no more than $p-1$. Consider a target set
$\mathcal{S}=\{o'_1,\ldots,o'_p\}$ s.t.
$\textsf{app}_{\mathcal{S}}(o'_1)\leq\ldots\leq\textsf{app}_{\mathcal{S}}(o'_p)$.
Then $k>|\bigcup_{1\leq t\leq k}\mathcal{S}^*_{t}|$ for any $k\geq
1$ iff $k>|\bigcup_{1\leq t\leq k}\mathcal{S}^*_{t}|$ for any
$p\geq k\geq 1$. Let $\mathcal{S}'=\mathcal{S}\setminus\{o'_p\}$
and
$\rho_{\mathcal{S}'}=\langle\mathcal{O}'_1,\mathcal{S}_1,\mathcal{O}_1\rangle,\langle\mathcal{O}'_2,\mathcal{S}_2,\mathcal{O}_2\rangle,\ldots$.
\begin{itemize}
\item If $k>|\bigcup_{1\leq t\leq k}\mathcal{S}^*_{t}|$ for any
$p\geq k\geq 1$ then:
\begin{enumerate}
\item $\textsf{app}_{\mathcal{S}}(o'_p)>|\bigcup_{1\leq t\leq
\textsf{app}_{\mathcal{S}}(o'_p)}\mathcal{S}^*_{t}|=p$. So
$o'_p\not\in\bigcup_{t\geq
1}^p(\mathcal{S}^*_t\cup\mathcal{O}^*_t)$. From Lemma 3, we have
$o'_p\not\in\bigcup_{t\geq
1}^p(\mathcal{S}_t\cup\mathcal{O}_t)\subseteq\bigcup_{t\geq
1}^p(\mathcal{S}^*_t\cup\mathcal{O}^*_t)$ and $k>|\bigcup_{1\leq
t\leq k}\mathcal{S}^*_{t}|\geq|\bigcup_{1\leq t\leq
k}\mathcal{S}_{t}|$ for any $p\geq k\geq 1$.

\item According to item 1 and the assumption, there exists a
successful strategy $\tau'$ for
$\langle\langle\succ_2,\ldots,\succ_n\rangle,\mathcal{S}'\rangle$
starting by reporting the objects in $\mathcal{S}'$.

\item According to items 1 and 2, if at each stage $k<p$, 1
reports the object specified by $\tau'$, then $o'_p$ is available
and not reported by any $i\in\{2,\ldots,n\}$ at stage $p$. Let
$\tau$ be a well--defined strategy reporting the object specified
by $\tau'$ at any stage $k<p$, and reporting $o'_p$ at stage $p$.
It is easy to find that $\tau$ is successful for $M$.
\end{enumerate}

\item If there exists some $p\geq k\geq 1$ s.t.
$k\leq|\bigcup_{1\leq t\leq k}\mathcal{S}^*_{t}|$, then there must
be some $i\in\{2,\ldots,n\}$ reporting some $o\in\bigcup_{1\leq
t\leq k}\mathcal{S}^*_{t}$ at some stage $k'\leq k$. In this case,
there is no successful strategy for $M$.
\end{itemize}
So the statement holds for any target set whose size is $p$.
$\quad\square$
\end{small}

We develop Algorithm 1 (in which the set of objects
$\mathcal{O}=\{o_1,\ldots,o_m\}$ and the set of agents
$\mathcal{N}=\{1,\ldots,n\}$ are supposed to be global variables)
to find successful strategies. The soundness and and completeness
of the algorithm is from the proof of Theorem 1. It is not hard to
find that Algorithm 1 always terminates and is polynomial in $m$
and $n$.\\
\\
\begin{small}
\begin{tabular}{rl}
\hline\noalign{\smallskip}
& \textbf{Algorithm 1:} Finding a successful strategy $\tau$ for $M$\\
\hline\noalign{\smallskip}
& \textbf{input:} a manipulation
problem
$M=\langle\langle\succ_2,\ldots,\succ_n\rangle,\mathcal{S}\rangle$\\
& \textbf{output:} a successful strategy
$\tau$ for $M$ if it exists,\\
& $\qquad\quad\;\;$otherwise \textsf{failure};\\
1. & $\mathcal{O'}\leftarrow\mathcal{O}$, $\mathcal{S'}\leftarrow\mathcal{S}$, $size\leftarrow 0$, $\tau\leftarrow\varepsilon$, $\tau'\leftarrow\varepsilon$,\\
& $\mathcal{N'}\leftarrow\{2,\ldots,n\}$, $k\leftarrow 1;\quad$  /* Initialization */\\
2. & $\textbf{while}(\mathcal{O'}\neq\emptyset)$\\
3. & $\quad\mathcal{S}^*\leftarrow\bigcup_{i\in\mathcal{N'}}\textsc{Better}_i(\mathcal{S'},\mathcal{O'}\setminus\mathcal{S'})$;\\
4. & $\quad size\leftarrow size+|\mathcal{S}^*|$;\\
5. & $\quad\textbf{if}$ $size\geq k$ $\textbf{then}$ $\textbf{return}$ $\textsf{failure}$;\\
6. & $\quad\textbf{for}$ all $o\in\mathcal{S}^*$\\
7. & $\qquad \tau\leftarrow\tau\bullet o;\quad$ /* $\bullet$ denotes connection*/\\
8. &
$\quad\mathcal{O}^*\leftarrow\{o\in\mathcal{O}'\setminus\mathcal{S}'|(\exists
i\in\mathcal{N}')o=\textsc{Best}_i(\mathcal{O}'\setminus\mathcal{S}')\}$;\\
9. & $\quad\textbf{if}$ $k>|\mathcal{S}|$ and $\mathcal{O}^*\neq\emptyset$\\
10. & $\qquad$randomly pick an object $o$ from $\mathcal{O}^*$;\\
11. & $\qquad \tau'\leftarrow\tau'\bullet o$; /*completed so as to be well--defined*/\\
12. & $\quad\mathcal{O}'\leftarrow\mathcal{O}'\setminus(\mathcal{S}^*\cup\mathcal{O}^*)$, $\mathcal{S}'\leftarrow\mathcal{S}'\setminus\mathcal{S}^*$, $k\leftarrow k+1$;\\
13. & $\textbf{return}$ $\tau\bullet\tau'$;\\
\noalign{\smallskip} \hline
\end{tabular}
\end{small}

We say a well-defined strategy $\tau$ is \emph{optimal} (in the
sense of pessimism) if it maximizes 1's benefit under the
assumption that 1 can not win any lottery. In fact, it is not hard
to find that if agent 1's scoring function is $g_L$ then she can
find an optimal strategy (in the sense of pessimism) in polynomial
time by applying Algorithm 2.\\
\\
\begin{small}
\begin{tabular}{rl}
\hline\noalign{\smallskip}
& \textbf{Algorithm 2:} Finding an optimal strategy\\
\hline\noalign{\smallskip} & \textbf{input:} a profile $R=\langle\succ_1,\ldots,\succ_n\rangle$\\
& \textbf{output:} an optimal strategy $\tau$ in the sense of
pessimism\\
1. & $\mathcal{O'}\leftarrow\mathcal{O}$, $\mathcal{S}\leftarrow\emptyset;\quad$  /* Initialization */\\
2. & $\tau\leftarrow\textbf{Algorithm 1}(\langle\succ_2,\ldots,\succ_n\rangle,\emptyset)$;\\
3. & $\textbf{while}(\mathcal{O'}\neq\emptyset)$\\
4. & $\quad
o\leftarrow\textsc{Best}_1(\mathcal{O}')$, $\mathcal{O}'\leftarrow\mathcal{O}'\setminus\{o\}$;\\
5. & $\quad \tau'\leftarrow\textbf{Algorithm 1}(\langle\succ_2,\ldots,\succ_n\rangle,\mathcal{S}\cup\{o\})$;\\
6. & $\quad\textbf{if}$ $\tau'\neq\textsf{failure}$\\
7. & $\qquad \tau\leftarrow\tau',\mathcal{S}\leftarrow\mathcal{S}\cup\{o\};$\\
8. & $\textbf{return}$ $\tau$;\\
\noalign{\smallskip} \hline
\end{tabular}
\end{small}
\\
\\
Let's run Algorithm 2 on the profile $R$ given in Example 1. Then
$\{o_2\}$, i.e., the best set of objects that 1 can manage to get
is found, and a successful strategy for the set (i.e, $o_2,o_3$ or
$o_2,o_5$) is returned. We conjecture that under the Borda scoring
function $g_B$, the problem of finding an optimal strategy is
\textsf{NP}-hard, but we do not have a proof.

\section{Conclusion}
We have defined and studied a parallel elicitation-free protocol
for allocating indivisible goods. The protocol is parameterized by
a parallel policy (i.e., an agent selection policy), which can
consider the allocation history that had happened. We have
compared a special parallel policy (i.e., $\varpi_A$) with
sequential policies for small numbers of objects and agents with
respect to the three social welfare criteria considered in
\cite{Bouveret} and \cite{Kalinowski}. The results show that
$\varpi_A$ outperforms the optimal sequential policies in most
cases. We have also proved that an agent who knows the preferences
of other agents can find in polynomial time whether she has a
successful strategy for a target set; and that if the scoring
function of the manipulator is $g_L$, she could compute an optimal
strategy (in the sense of pessimism) in polynomial time.

There are several directions for future work. One direction would
be to prove the conjectures about the social welfare induced by
$\varpi_A$, and to design other parallel policies that can
outperform $\varpi_A$ in some social welfare criterion. Another
direction would be to find the missing complexity results for
manipulation under $g_B$, and to consider strategical issues under
the assumption that the manipulator believes any lottery is fair.
Furthermore, we plan to design an elicitation-free protocol for
allocating sharable goods \cite{Airiau}.

\section*{Acknowledgments}
This work is supported by the National Natural Science Foundation
of China under grant No.61070232, No.61105039, No.61272295, and
the Fundamental Research Funds for the Central Universities (Issue
Number WK0110000026). Many thanks to the anonymous reviewers for
their comments.

\bibliographystyle{named}
\bibliography{ijcai13}

\end{document}